 \documentclass[CM,Submssn]{degruyter-nocrelle}       

\title[A Theta lift representation for  the Kawazumi-Zhang  invariant]{A Theta lift representation for  the Kawazumi-Zhang and  Faltings invariants of genus-two Riemann surfaces}

\author{Boris}{Pioline}{B.~Pioline}{Geneva}

\contact{CERN, Theory Division, Case C01600, CH-1211 Geneva 23\\On leave from: 
LPTHE, CNRS UMR 7589 and Universit\'e Pierre et Marie Curie, \\ 4 place Jussieu, F-75005, Paris, France.}{boris.pioline@cern.ch}

\theoremstyle{plain}
 \newtheorem{theorem}{Theorem}

\theoremstyle{definition}

\theoremstyle{plain}
 \newtheorem{corollary}{Corollary}

\def\be{\begin{equation}}
\def\ee{\end{equation}}
\newcommand{\IR}{\mathbb{R}}

\newcommand{\IZ}{\mathbb{Z}}
\newcommand{\pa}{\partial}

\newcommand{\cF}{\mathcal{F}}
\newcommand{\cS}{\mathcal{S}}

\newcommand{\cK}{\mathcal{K}}
\newcommand{\cI}{\mathcal{I}}
\newcommand{\cM}{\mathcal{M}}

\newcommand{\cN}{\mathcal{N}}

\newcommand{\cR}{\mathcal{R}}

\newcommand{\cO}{\mathcal{O}}
\newcommand{\cH}{\mathcal{H}}

\newcommand{\cA}{\mathcal{A}}

\newcommand{\cD}{\mathcal{D}}

\newcommand{\I}{\mathrm{i}}
\newcommand{\de}{\mathrm{d}}
\DeclareMathOperator{\Li}{Li}
\renewcommand{\Im}{\operatorname{Im}}
\renewcommand{\Re}{\operatorname{Re}}

\begin{document}

\begin{abstract}
The Kawazumi-Zhang invariant $\varphi$ for compact genus-two Riemann surfaces was recently shown to be an eigenmode of the Laplacian on the Siegel upper half-plane, away from the separating degeneration divisor. Using this fact and the known behavior of $\varphi$ in the non-separating degeneration limit, it is shown that $\varphi$ is equal to the Theta lift of the unique (up to normalization) weak Jacobi form of weight $-2$. This identification provides the complete Fourier-Jacobi expansion of $\varphi$ near the non-separating node,  gives full control on the asymptotics of $\varphi$ in the various degeneration limits, and provides an efficient numerical procedure to evaluate $\varphi$ to arbitrary accuracy. It also reveals a mock-type holomorphic Siegel modular form of weight $-2$ underlying $\varphi$. From the general relation between the Faltings invariant, the Kawazumi-Zhang invariant and the discriminant for hyperelliptic Riemann surfaces, a Theta lift representation for the Faltings invariant in genus two readily follows.\end{abstract}


\section{Introduction}\label{sec:intro}

The Kawazumi-Zhang invariant, introduced in \cite{Kawazumi,zbMATH05661751}, is a real-valued function 
$\varphi(\Sigma)$ on the moduli space $\cM_h$ of compact Riemann surfaces $\Sigma$ of 
genus $h\geq 1$.  One way of defining it is through the spectrum of the Laplacian $\Delta_\Sigma$ with respect to the Arakelov metric on $\Sigma$,
\be
\varphi(\Sigma) = \sum_{\ell>0} \frac{2}{\lambda_\ell} \sum_{m,n=1}^{h} 
\left| \int_\Sigma\, \phi_\ell \, \omega_m \, \bar\omega_n \right|^2
\ee
where $(\omega_1,\dots,\omega_h)$ is an orthonormal basis of holomorphic differentials on $\Sigma$,
$0=\lambda_0<\lambda_1\leq$  $\lambda_2\leq \dots$ are the eigenvalues of $\Delta_\Sigma$, and
 $\phi_\ell$ a corresponding orthonormal basis of real square-integrable eigenmodes. For genus 
$2$, $\varphi(\Sigma)\equiv\varphi(\Omega)$ is a function of the period matrix $\Omega$, and defines a real-analytic modular function on the Siegel upper half-plane $\cH_2$, away from the separating degeneration divisor. The Kawazumi-Zhang invariant is a close cousin \cite{zbMATH06139356} of the Faltings invariant $\delta_F(\Sigma)$ \cite{zbMATH03891504}, which plays an important role in arithmetic geometry. Its asymptotic behavior near the boundaries of the moduli space $\cM_h$ in arbitrary genus was investigated in \cite{MR1105425,zbMATH06355718,zbMATH05725877}.

While Faltings' invariant made an appearance in studies of bosonisation in conformal field theory \cite{AlvarezGaume:1987vm}, the genus-two Kawazumi-Zhang invariant has entered the physics literature in a recent analysis of the low energy expansion of the two-loop four-graviton amplitude in superstring theory   \cite{zbMATH06339620}:  the leading $D^4 \cR^4$ interaction is proportional to the Weil-Petersson volume of the moduli space $\cM_2$, while the 
next-to-leading  $D^6 \cR^4$ interaction is proportional
to the integral of $\varphi$ times the same Weil-Petersson 
volume form on $\cM_2$. With hindsight from various physics conjectures, 
it was proven in \cite{D'Hoker:2014gfa} that $\varphi$ is an eigenmode of the Laplacian 
$\Delta_{Sp(4)}$ on the Siegel upper half-plane, up to a source term supported on the separating degeneration divisor,
\be
\label{kzlap}
\left[ \Delta_{Sp(4)} -5 \right]\, \varphi = -2\pi\,\det(\Im\Omega)\, \delta^{(2)}(v)\ ,
\ee
where $v$ is the off-diagonal element in the period matrix
\be
\label{omrvs}
\Omega =\begin{pmatrix} \rho &v \\ v & \sigma \end{pmatrix} =
\begin{pmatrix} \rho_1 &v_1 \\ v_1 & \sigma_1 \end{pmatrix} + \I
\begin{pmatrix} \rho_2 &v_2 \\ v_2 & \sigma_2 \end{pmatrix} \ .
\ee
As we shall see, this partial differential equation provides strong constraints on the asymptotic behavior at the boundaries of $\cM_2$. 

In the maximal degeneration limit, where all entries in the imaginary part  $\Omega_2$ of the period matrix $\Omega=\Omega_1+\I \Omega_2$ are scaled to infinity, it was shown in \cite{zbMATH05725877,D'Hoker:2014gfa} that
\be
\label{kzmaxdeg}
\varphi(\Omega) = \frac{\pi}{6}
\left[  L_1+L_2+L_3 - \frac{5\, L_1 L_2 L_3}{L_1 L_2+L_2 L_3+L_3 L_1} \right] + \cO(1/L_i^2)
\ee
where $0<L_3\leq L_1\leq L_2$ parametrize the imaginary part of $\Omega$ in the standard fundamental domain of the action of $GL(2,\IZ)$ on the space of $2\times 2$ positive definite real matrices,
\be
\label{eq:omegaL}
\Omega_2 = \begin{pmatrix} L_1+L_3 & L_3 \\ L_3 & L_2+L_3 \end{pmatrix}\ .
\ee
This parametrization is motivated by the connection to two-loop supergravity amplitudes, where the $L_i$'s play the role of Schwinger time parameters \cite{Green:2005ba,Green:2008bf,D'Hoker:2014gfa}. 
The leading term in \eqref{kzmaxdeg} is an exact solution of \eqref{kzlap} with no source term, which was one of the hints towards the differential equation \eqref{kzlap} in \cite{D'Hoker:2014gfa}.

In the minimal degeneration limit $\sigma\to\I\infty$, keeping the other entries of $\Omega$ fixed, one has instead \cite{MR1105425,zbMATH06355718}
\be
\label{kzmindeg}
 \varphi(\Omega) = 
  \frac{\pi}{6} t  +\varphi_0(\rho,u_1,u_2) + 
\cO(1/t)
\ee
where 
\be
\label{defphi0}
\varphi_0(\rho,u_1,u_2) = - \log\left[ e^{-\pi\rho_2 u_2^2}  \left | {\theta( \rho,\rho u_2-u_1) \over \eta (\rho)} \right | \right]\ .
\ee
Here, $v= \rho u_2 - u_1$ where $u_1, u_2$ are real, 
$t=\sigma_2-u_2^2 \rho_2$ is non-negative, $\eta(\rho)$ is Dedekind's eta function and $\theta(\rho,v)=\sum_{n\in\IZ} e^{\I\pi(n+\frac12)^2 \rho+2\pi \I (n+\frac12)(v+\tfrac12)}$ is Jacobi's theta series.
The first two terms in \eqref{kzmindeg} satisfy  \eqref{kzlap} up to terms of order $1/t$. Up to the order displayed, each term in the Laurent expansion around $t=\infty$ is  a real-analytic function of  $\rho,u_1,u_2$ invariant under the Jacobi subgroup $\Gamma_J=SL(2,\IZ)\ltimes \IZ^2\ltimes \IZ$ of the Siegel modular group $\Gamma=Sp(4,\IZ)$ (i. e.  
a Jacobi form of weight zero and index zero). As we shall see, this structure extends
to all orders in $1/t$.

Finally, in the separating degeneration limit $v\to 0$, keeping $\rho,\sigma$ fixed, one 
has~\cite{MR1105425,zbMATH06355718}
\be
\label{kzsepdeg}
\varphi(\Omega) = -\log \left| 2\pi v \, \eta^2(\rho)  \eta^2(\sigma) \right| +\cO(|v|^2 \log |v|)\ .
\ee
Eq. \eqref{kzsepdeg} is 
consistent with the differential equation \eqref{kzlap} at the order stated, with the logarithmic behavior at $v=0$ being responsible for the delta-function source term. In \cite{D'Hoker:2014gfa} it was shown using the properties above that the average value of $\varphi$ on $\cM_2$ with respect to the Weil-Petersson volume form is equal to 3/2, verifying a prediction from S-duality in superstring theory.

Our goal in this work is to determine the complete 
asymptotics  of the invariant $\varphi(\Omega)$ in the degeneration limits \eqref{kzmaxdeg} and \eqref{kzmindeg}, and more generally, obtain the  complete Fourier expansion with respect to $\Omega_1$. 
To this aim, in \S\ref{sec_kztheta} we shall construct a real-analytic Siegel modular form $\tilde\varphi(\Omega)$ on $\cM_2$  that satisfies \eqref{kzlap}, \eqref{kzmaxdeg} and \eqref{kzmindeg}. Since  $\varphi-\tilde\varphi$ is square integrable on $\cM_2$ and an eigenmode of $\Delta_{Sp(4)}$ with non-negative eigenvalue, it must therefore vanish (Theorem 1 in \S\ref{sec_thm}). 
$\tilde\varphi$ is constructed as the Theta lift  of 
the unique weight $-2$ weak Jacobi form $\theta^2/\eta^6$ (see Eq. \eqref{phithetalift}). 
This parallels the construction
of the log-norm of the Igusa cusp form $\Psi_{10}$ as the Theta lift of the unique weight 0 weak Jacobi 
form (also known as the elliptic genus of K3) due to Kawai \cite{Kawai:1995hy}, 
which we review in \S\ref{sec_psi10}.  Since singular Theta lifts were studied extensively
in \cite{0919.11036,1004.11021}, we refer to these works for issues of convergence. For the convenience of the reader however, we shall rederive the Fourier expansions at a physicist's level of rigor. Using the relation between the Faltings invariant,
the Kawazumi-Zhang invariant and $\Psi_{10}$ established in \cite{zbMATH06139356}, a Theta lift
representation for the Faltings invariant is readily obtained (Corollary 3 in \S\ref{sec_thm}). 

The Theta lift representation of $\varphi$ has several interesting consequences, considered in \S\ref{sec_thm} and \S\ref{sec4}. First, it gives complete control over the asymptotics in the various degeneration limits, and provides an efficient numerical procedure to evaluate $\varphi$ to arbitrary accuracy. This is likely to have useful applications in Arakelov geometry.  Second, it reveals a `holomorphic prepotential' $F_1(\Omega)$ which generates $\varphi$ through the action of (the real part of) the Siegel-Maass raising operator (Eq. \eqref{phiBoxF}). $F_1$  transforms non-homogeneously under $Sp(4,\IZ)$, giving an explicit example of a mock-type Siegel modular form. Third, it implies that $\varphi$ is an eigenmode of an invariant quartic differential operator (Eq. \eqref{quarticphi}). It would be interesting to understand whether the differential equations \eqref{kzlap} and \eqref{quarticphi} can be generalized to higher genus.

From the physics point of view, the results obtained here will be key for checking S-duality predictions for  $D^6\cR^4$ couplings in string theory  \cite{PiolineRusso-to-appear}. In a different vein, it is worth noting that 
the same type of prepotential $F_1$ appears in the physics literature when computing one-loop
corrections to the holomorphic prepotential  in heterotic vacua with $\cN=2$ supersymmetry \cite{Mayr:1993mq,Antoniadis:1995ct,Harvey:1995fq}. In that context, $F_1$ encodes a subset of the Gromov-Witten invariants in the dual type IIA string theory compactified on a suitable K3-fibered Calabi-Yau threefold. This analogy suggests that the product of the moduli space of genus-two Riemann surfaces times the  Poincar\'e upper half-plane $\cH_1$ (parametrizing the size  $s$ of the base of the K3-fibration) may carry some canonical special K\"ahler metric derived from a prepotential 
$F(s,\rho,v,\sigma)=s(\rho\sigma-v^2) + F_1 + \cO(e^{-s})$, where $F_1$ is the holomorphic prepotential underlying the Kawazumi-Zhang invariant. It would be very interesting to find a string theory compactification whose moduli space carries this putative metric, and compute the $\cO(e^{-s})$ corrections using mirror symmetry techniques.

\section{Refined degeneration formulae\label{sec1}}

In this section, we shall attempt to improve the accuracy of the asymptotic expansions \eqref{kzmaxdeg},  \eqref{kzmindeg} and  \eqref{kzsepdeg} by requiring consistency with the Laplace equation \eqref{kzlap} and invariance under the Jacobi group $\Gamma_J$. This section is heuristic, and the proof that $\varphi$ actually satisfies these improved asymptotic expansions is deferred to \S\ref{sec2}. This attempt is inspired by a  study of two-loop amplitudes  in superstring theory \cite{PiolineRusso-to-appear}, and in turn, in combination with insights gained from a study of generalized Borcherds lifts \cite{Angelantonj:2012gw,Angelantonj:2015rxa,AFP5-to-appear}, inspired the educated guess considered in \S\ref{sec_kztheta}. The  reader uninterested by the source of this guess can safely skip to \S\ref{sec2}.

Starting with the minimal non-separating degeneration, we observe that the expansion \eqref{kzmindeg} can be strengthened, consistently with the Laplace equation \eqref{kzlap} to exponential accuracy, to 
\be
\label{kzmindeg2}
 \varphi(\Omega) =  \frac{\pi}{6} t + \varphi_0 + \frac{\varphi_1}{t} + \cO(e^{-t})\ ,
\ee
where $\varphi_1$ is a function of $\rho, u_1,u_2$ to be determined. Indeed, decomposing the Laplace operator into 
\be
\Delta_{Sp(4)} = \Delta_t + \Delta_{\rho} + t\, \Delta_u + \Delta_{\sigma_1}
\ee
where
\be
\begin{split}
\Delta_t =&  t^2 \pa^2_t - t\pa_t\ ,\quad \Delta_{\rho} =  \rho_2^2 ( \pa_{\rho_1}^2+\pa_{\rho_2}^2)\ ,\quad
\Delta_u = \frac{1}{2\rho_2} | \rho \,\partial_{u_1} + \partial_{u_2}|^2\ ,\quad 
\\
\Delta_{\sigma_1}=& - (t+\rho_2 u_2^2) \pa_{\sigma_1}^2 + (2 t \rho_2 u_2 \pa_{u_1}-2\rho_2^2 u_2^2 \pa_{\rho_1})
\pa_{\sigma_1}\ ,
\end{split}
\ee
and using the fact that $\varphi_0$, defined in \eqref{defphi0}, satisfies
\be
\Delta_{\rho}\, \varphi_0 = 0\ ,\quad \Delta_u\, \varphi_0 = \pi  \ ,
\ee
we see that the Laplace equation \eqref{kzlap} is satisfied at order $\cO(e^{-t})$ provided
$\varphi_1$ satisfies
\be
\label{phi1eq}
\begin{split}
\Delta_{\rho} \,\varphi_1 = 2 \varphi_1\ ,\quad \Delta_u\, \varphi_1 = 5 \varphi_0\ .
\end{split}
\ee
Invariance of $\varphi$ under $\Gamma$ requires that $\varphi_1$ be a 
real-analytic Jacobi form of zero weight and zero index.  
On the other hand, the maximal degeneration limit \eqref{kzmaxdeg}
requires that, in the limit $\rho_2\to\infty$ keeping $\rho_1,u_1,u_2$ fixed (with $u_2\in[0,1]$),
\be
\label{phi01lim0}
\varphi_0 \sim \frac{\pi}{6}\rho_2(1-6u_2+6u_2^2)\ ,\quad 
\varphi_1 \sim \frac{5\pi}{6} \rho_2^2 u_2^2(u_2-1)^2\ .
\ee
The first equation is of course satisfied by \eqref{defphi0}.
It is suggestive to rewrite these limits   in terms of the Bernoulli polynomials $B_2(x)=x^2-x+\tfrac16$, $B_4(x)=x^2(x-1)^2-\tfrac{1}{30}$:
\be
\label{phi01lim1}
\varphi_0 \sim \pi\rho_2 B_2(u_2)\ ,\quad \varphi_1 \sim \frac{5\pi}{6} \rho_2^2 \left(B_4(u_2)+\tfrac{1}{30}\right)
\ee
A solution to \eqref{phi1eq} obeying these boundary conditions can be obtained 
as a linear combination
\be
\label{phi1D22}
\varphi_1 = \frac{5}{16\pi^2\rho_2}\, \cD_{2,2}(\rho;v) +\frac{5}{2\pi}\, E^{\star}(2;\rho)
\ee
of the standard non-holomorphic Eisenstein series 
\be
E^\star(s;\rho) = \frac12 \pi^{-s} \Gamma(s) \sum_{(m,n)\neq(0,0)} \left[\frac{\rho_2}{|m\rho+n|^2}\right]^s
\ee
and 
the Kronecker-Eisenstein series introduced in \cite{zbMATH04144378} 
\begin{equation}
\label{KronEisZ}
\cD_{a,b} (\rho ; v ) \equiv \frac{(2\I \rho_2)^{a+b-1}}{2\pi\I}\, \sum_{(m,n)\neq(0,0)}\, \frac{e^{2\pi\I (n\, u_2 + m \, u_1 )}}{(m\rho + n)^a (m\bar\rho+n)^b}\ ,
\end{equation}
where   $a,b$ are non-negative integers.
$\cD_{a,b} (\rho ; v )$ is a real-analytic Jacobi modular form of weight $(1-b,1-a)$ and zero index,
with Fourier expansion 
\begin{equation}
\label{FourierDabqx}
\cD_{a,b} (\rho;v) = \sum_{m=0}^\infty D_{a,b} (q^m \, x) + (-1)^{a+b} \sum_{m=1}^\infty D_{a,b} (q^m\, x^{-1}) + \frac{ (4\pi\rho_2)^{a+b-1}}{(a+b)!}\, B_{a+b} \left( u_2 \right)\,,
\end{equation}
where $x= e^{2\pi\I v}=e^{2\pi\I(u_2\rho-u_1)}$, $q=e^{2i\pi \rho}$, $B_\alpha (x)$ are the Bernoulli polynomials, and $D_{a,b}(x)$ are the  Bloch-Wigner-Ramakrishnan single-valued polylogarithms \cite{zbMATH04144378},
\begin{equation}\label{eq:Dabdef}
\begin{split}
D_{a,b} (x) =& (-1)^{a-1} \sum_{k=a}^{a+b-1} 2^{a+b-1-k} \, \left( {k-1 \atop a-1}\right) 
\frac{(-\log\, |x|)^{a+b-1-k}}{(a+b-1-k)!}\, {\rm Li}_{k} (x)\\
& + (-1)^{b-1} \sum_{k=b}^{a+b-1} 2^{a+b-1-k} \, \left( {k-1 \atop b-1}\right) 
\frac{(-\log\, |x|)^{a+b-1-k}}{(a+b-1-k)!}\, \overline{{\rm Li}_{k} (x)}\ .
\end{split}
\end{equation}
It is easy to check that the differential equations \eqref{phi1eq} are obeyed, by checking the action on the seed of the Poincar\'e series (i.e. setting $m=0,n=1$) and using the second Kronecker limit formula, which states
\be
\label{phi0D11}
\varphi_0 = \frac12\, \cD_{1,1}(\rho;v) \ .
\ee
 Moreover, the equality \eqref{phi1D22} predicts an additional subleading term in \eqref{phi01lim1},
\be
\label{E2starexp}
\varphi_1 = \frac{5\pi}{6} \rho_2^2 B_4(u_2)
+ \frac{\pi}{36} \, \rho_2^2 + \frac{5\zeta(3)}{4\pi^2} \, \rho_2^{-1} + \cO(e^{-2\pi\rho_2})
\ee
The third  term in \eqref{E2starexp} requires a subleading correction 
to the maximal degeneration limit \eqref{kzmaxdeg},
\be
\label{kzmaxdeg2}
\varphi(\Omega) = \frac{\pi}{6}
\left[  L_1+L_2+L_3 - \frac{5\, L_1 L_2 L_3}{L_1 L_2+L_2 L_3+L_3 L_1} \right] + 
 \frac{5\zeta(3)}{4\pi^2 (L_1 L_2+L_2 L_3+ L_3 L_1)}+ \dots 
\ee
The additional term is an exact solution of the Laplace equation \eqref{kzlap}. 

With  the hindsight gained from
a study of generalized Borcherds lifts \cite{AFP5-to-appear}, the estimates \eqref{kzmindeg2} and \eqref{kzmaxdeg2}, if true, strongly suggest that $\varphi(\Omega)$ is the Theta lift of an almost, weakly holomorphic Jacobi form of weight $-1/2$ and depth $1$, motivating the educated guess in \S\ref{sec_kztheta}. We shall prove in \S\ref{sec_thm}
that this estimates  do in fact hold with exponential accuracy.

\section{The Kawazumi-Zhang invariant as a Theta lift\label{sec2}}

In this section, using a suitable Theta lift, we construct a real-analytic Siegel 
modular form $\tilde\varphi$  that satisfies the 
differential equation \eqref{kzlap} and asymptotic behaviors \eqref{kzmaxdeg},  \eqref{kzmindeg}, 
 \eqref{kzsepdeg} in the various degeneration limits -- hence must coincide with 
 $\varphi$. As a warm-up, we start by recalling the Theta lift representation of the log-norm of
 the discriminant of genus two Riemann surfaces, following \cite{Kawai:1995hy}.
 
\subsection{The Igusa cusp form $\Psi_{10}$ as a Theta lift\label{sec_psi10}}

Recall that the log-norm $\log||\Psi_{10}||$ $=\log[(\det\Omega_2)^5|\Psi_{10}|]$ of the weight 10 Igusa cusp form (normalized as $2^{-12}$ times the product of the squares of the ten Thetanullwerte) can be represented as a regularized modular integral\footnote{The formula \eqref{thetalift} was discovered in \cite{Kawai:1995hy} by computing threshold corrections to gauge couplings in heterotic string theory compactified on $K3\times T^2$. The variables 
$\rho,\sigma,v$ parametrize the complex structure, K\"ahler class and holonomies of a $U(1)$ connection on the torus $T^2$, while the integers $m_i, n^i$ correspond to the momentum and winding numbers, and $b$ is the electric charge.\label{footKawai}}
\be
\label{thetalift}
\begin{split}
\log||\Psi_{10}||(\Omega)=& -\frac14 \int_{\cF_1} \frac{\de^2\tau}{\tau_2^2}\,
\left[ \Gamma^{\rm even}_{3,2}(\Omega;\tau)\, h_0(\tau) +  \Gamma^{\rm odd}_{3,2}(\Omega;\tau)\, h_1(\tau)
- 20\, \tau_2 \right] \\
& - 5 \log \left( \frac{8\pi}{3\sqrt3} e^{1-\gamma_E}\right)\ ,
\end{split}
\ee
where $\cF_1=\{\tau \in \cH_1, |\tau|>1, -\tfrac12<\tau_1\leq \tfrac12\}$ is the standard fundamental domain for the action of $SL(2,\IZ)$ on $\cH_1$, 
$\de^2\tau=\de \tau_1 \de \tau_2$, $\gamma_E$ is the Euler-Mascheroni constant, and $\Gamma^{\rm even \vert odd}_{3,2}$
are partition functions for even (shifted) lattices of signature (3,2), or Siegel-Narain theta series, 
\begin{equation}
\label{defGamma32}
\begin{split}
\Gamma^{\rm even\vert odd}_{3,2}(\Omega;\tau)=& \tau_2\, \sum_{\substack{
(m_1,m_2,n^1,n^2)\in \IZ^4\\ b\in 2\IZ \vert 2\IZ+1}} 
q^{\frac14 p_L^2}\,
\bar q^{\frac14 p_R^2}\\
p_{\rm R}^2 =& \frac{|m_2 - \rho\, m_1 + \sigma n^1+ (
\rho\sigma- v^2)\, n^2 
- b\, v|^2}{\rho_2\, \sigma_2-v_2^2}   ,\\
p_{\rm L}^2 =& p_R^2 + 4 m_i n^i + b^2  \, .
\end{split}
\end{equation}
For $v\to 0$, $\Gamma^{\rm even}_{3,2}\to \Gamma_{2,2}\,\theta_3(2\tau), 
\Gamma^{\rm odd}_{3,2}\to \Gamma_{2,2}\,\theta_2(2\tau)$ where $\Gamma_{2,2}$ is the partition
function of the even self-dual lattice with signature $(2,2)$,
\be
 \Gamma_{2,2}(\rho,\sigma;\tau) = \tau_2\, \sum_{(m_1,m_2,n^1,n^2)\in \IZ^4} q^{\frac14 p_L^2}\,
\bar q^{\frac14 p_R^2}\vert_{b=v=0}\ .
\ee 
Thus, $\Gamma^{\rm even\vert odd}_{3,2}$
are modular forms of weight $1/2$ under $\Gamma_0(4)$.
Under general modular transformations of $\tau$,
\be
\begin{split}
\Gamma^{\rm even}_{3,2}(\tau+1)=&\Gamma^{\rm even}_{3,2}(\tau)\ ,\quad 
\Gamma^{\rm odd}_{3,2}(\tau+1) = \I \,\Gamma^{\rm odd}_{3,2}(\tau)
\\
\Gamma^{\rm even}_{3,2}(-1/\tau) =& \frac{1-\I}{2}\, \tau^{1/2}\, [\Gamma^{\rm even}_{3,2}(\tau) + \Gamma^{\rm odd}_{3,2}(\tau)]\ ,\quad \\
\Gamma^{\rm odd}_{3,2}(-1/\tau) =& \frac{1-\I}{2}\, \tau^{1/2}\, [\Gamma^{\rm even}_{3,2}(\tau) - \Gamma^{\rm odd}_{3,2}(\tau)]\ .
\end{split}
\ee
The integers $m_1,m_2,n^1,n^2,b$ can be fit into  an antisymmetric traceless matrix
\be
\begin{pmatrix}
0 & -m_2 & b/2 & n^1 \\
m_2 & 0 & m_1 & -b/2 \\
-b/2 & -m_1 & 0 &  -n^2 \\
-n^1 & b/2 & n^2 & 0
\end{pmatrix}\ ,
\ee
with Pfaffian proportional to $p_L^2-p_R^2$, which transforms by conjugation under $Sp(4,\IZ)$.This makes it clear that $Sp(4,\IZ)$ transformations preserve the parity of $b$.
Thus, both $\Gamma^{\rm even\vert odd}_{3,2}(\Omega;\tau)$ are 
Siegel modular functions in the variable $\Omega$, and so then is the result of the modular integral 
 \eqref{thetalift}. On the other hand, 
 $h_0, h_1$ are the coefficients of the theta series decomposition of the elliptic genus of $K3$,
\be
\begin{split}
\chi_{K3}(\tau,z) =& h_0(\tau) \, \theta_{3}(2\tau,2z) +h_1(\tau) \, \theta_{2}(2\tau,2z) \ ,\\
h_0(\tau)=&24 \frac{\theta_3(2\tau)}{\theta_3^2(\tau)} - 2 \frac{\theta_4^4(\tau)-\theta_2^4(\tau)}{\eta^6(\tau)} \theta_2(2\tau)
=20 + 216 q+1616 q^2+ \dots \\
h_1(\tau)=&24 \frac{\theta_2(2\tau)}{\theta_3^2(\tau)} + 2 \frac{\theta_4^4(\tau)-\theta_2^4(\tau)}{\eta^6(\tau)} \theta_3(2\tau)
= q^{-1/4} ( 2 - 128 q - 1026 q^2 + \dots)
\end{split}
\ee
They are modular forms of $\Gamma_0(4)$ with weight $-1/2$. In terms of the standard  
generators $X_2(2\tau)=$ $E_2(2\tau)-2 E_2(4\tau)$, $\theta_2^4(2\tau)$ of the ring
of $\Gamma_0(4)$ modular forms of even weight, 
\be
\label{defhi}
\begin{split}
h_0 =& \frac{1}{\theta_3(2\tau)\, \Delta_6}\left[
\frac{3}{16} \theta_2^{12}(2\tau) - X_2(2\tau) \theta_2^8(2\tau) + \frac54 [X_2(2\tau)]^2\, \theta^4_2(2\tau) \right] \\
h_1 =& \frac{1}{\theta_2(2\tau)\, \Delta_6}\left[
-\frac{9}{16} \theta_2^{12}(2\tau) + X_2(2\tau) \theta_2^8(2\tau) + \frac14 [X_2(2\tau)]^2\, \theta^4_2(2\tau) \right]
\end{split}
\ee
where $\Delta_6 = [\eta(2\tau)]^{12}$ is a cusp form of weight 6.  Under general modular transformations,
\be
\label{h012sl2}
\begin{split}
h_0(\tau+1)=&h_0(\tau)\ ,\quad h_1(\tau+1) = -\I h_1(\tau)\ ,
\\
h_0(-1/\tau) =& \frac{1+\I}{2}\, \tau^{-1/2}\, [h_0(\tau) + h_1(\tau)]\ ,\quad \\
h_1(-1/\tau) =& \frac{1+\I}{2}\, \tau^{-1/2}\, [h_0(\tau) - h_1(\tau)]\ ,\quad 
\end{split}
\ee
so that the integrand of \eqref{thetalift} is (except for the last term in the bracket, proportional to $\tau_2$) invariant under the full modular group $SL(2,\IZ)$. It follows from \eqref{h012sl2} 
(see e.g. the proof of Thm 5.4 in \cite{MR781735}) that
\be
\begin{split}
h(\tau) = & h_0(4\tau)+h_1(4\tau) \\
 = &  -\frac1{8 \theta_3(2\tau) \Delta_6}\left(
-3\theta_2^{12}(2\tau) + 4 \theta_2^8(2\tau) X_2(2\tau) +12  \theta_2^4(2\tau)  X_2^2(2\tau)
-16 X_2^3(2\tau) \right)\\
= &\sum_{m\geq -1} c(m) q^m = 2 q^{-1} + 20  - 128 q^3 + 216 q^4 - 1026 q^7+\dots
\end{split}
\ee
is modular form of weight $-\frac12$ under $\Gamma_0(4)$ in Kohnen's plus space ({\it i.e.}
the $m$-th Fourier coefficient $c(m)$ of $h(\tau)$ vanishes unless $m=0,3 \mod 4)$. 
$h(\tau)$ has a simple pole at $\tau=\I\infty$ and is regular at $\tau=0$ and $\tau=\frac12$.
The constant term in $h(\tau)$ makes it necessary to subtract by hand the term proportional to $\tau_2$ in \eqref{thetalift}, in order for the integral to converge.\footnote{Alternatively, following \cite{0919.11036} one could truncate the integration domain to $\cF_1^\Lambda=\cF_1\cap\{\tau_2<\Lambda\}$, insert a Kronecker regulating factor $\tau_2^s$ in the integrand, 
take the limit $\Lambda\to\infty$ for fixed
$s$ with $\Re(s)$ sufficiently large, analytically continue in $s$ and extract the constant term in the Laurent expansion at $s=0$. The two prescriptions can be shown to agree up to an additive constant.} 

Using the differential equation satisfied by the lattice partition function,
\be
\label{DelNarain}
\left[ \Delta_{Sp(4)} - 4\Delta_{SL(2),1/2} +1 \right]\, \Gamma^{\rm even| odd}_{3,2} = 0\ ,
\ee
where $\Delta_{SL(2),w}=4\tau_2^2 \partial_{\bar\tau} ( \partial_\tau + \frac{w}{2\I\tau_2}) + w$
is the Laplacian acting on modular forms of weight $w$, 
 one sees that $\log||\Psi_{10}||$ is a real-analytic quasi-harmonic function on the Siegel upper half-plane, up to a  delta function source term supported  on the separating divisor,
\be
\label{DelPsi10}
\Delta_{Sp(4)} \, \log||\Psi_{10}|| = -15 + 4\pi\, \delta^{(2)}(v)\ .
\ee
Indeed, as $v\to 0$, the integrand becomes
\be
\Gamma_{2,2} \left[ \theta_3(2\tau)\, h_0(\tau) + \theta_2(2\tau)\, h_1(\tau) \right] - 20 \tau_2
= 24 \Gamma_{2,2} - 20 \tau_2 \stackrel{\tau_2\to\infty}{\sim} 4\tau_2\ ,
\ee
which leads to a logarithmic divergence. Keeping $v$ small but non zero, and retaining the contributions from $m_i=n^i=0, b=\pm 1$, we have
\be
\log||\Psi_{10}|| \sim - \int_1^{\infty} \frac{\de\tau_2}{\tau_2} e^{-\frac{\pi\tau_2|v|^2}{\rho_2\sigma_2-v_2^2}}= - \Gamma\left(0,\pi z\right)\ ,
\ee
where $z=\frac{|v|^2}{\rho_2\sigma_2-v_2^2}$.
Using the fact that the incomplete Gamma function $\Gamma(0,\pi z)$ behaves as $-\log (\pi z)+
$ analytic as $z\to 0$, and the result from  \cite{Dixon:1990pc}
\be
\label{DKL}
\int_{\cF_1} \frac{\de^2\tau}{\tau_2^2} (\Gamma_{2,2}-\tau_2)=-\log
\left[ \frac{8\pi e^{1-\gamma}}{3\sqrt3} \rho_2 \sigma_2 |\eta(\rho) \eta(\sigma)|^4\right]\ ,
\ee
 we find
\be
\label{deglogpsi10}
\log||\Psi_{10}|| = \log |\rho_2^5 \sigma_2^5 v^2 \eta^{24}(\rho) \eta^{24}(\sigma)|+\cO(|v|^2)\ ,
\ee
where the omitted terms vanish analytically as $v\to 0$.

Evaluating the modular integral by the standard unfolding method \cite{Harvey:1995fq,0919.11036}, one arrives at 
\be
\label{logPsiprod}
\begin{split}
\log||\Psi_{10}||(\Omega) = & -2\pi(\rho_2+\sigma_2-v_2) + 5 \log \det \Omega_2 \\
& 
- \Re\Big[ \sum_{(k,\ell,b)>0} c(4k\ell-b^2) \log\left(1-e^{2\pi \I(k\sigma+\ell \rho+b v)}\right) \Big]\ ,
\end{split}
\ee
where $(k,\ell,b)>0$ stands for $\{ (k>0, \ell\geq 0) \, \operatorname{or}\, (k\geq 0, \ell> 0) , b\in\IZ\} \cup 
\{k=\ell=0,b>0\}$, and $\Omega$ is assumed to be such that $k\sigma_2+\ell\rho_2+b v_2>0$ for all $(k,\ell,b)>0$ \cite[Eq. (20)]{Kawai:1995hy}. Eq. \eqref{logPsiprod} is consistent with the Gritsenko-Nikulin product formula \cite{MR1428063}
\be
\label{Psiprod}
\Psi_{10}(\Omega)=e^{2\pi \I(\rho+\sigma-v)} \prod_{(k,\ell,b)>0} (1-e^{2\pi \I(k\sigma+\ell \rho+b v)})^{c(4k\ell-b^2)} \ .
\ee

\subsection{An educated guess\label{sec_kztheta}}

Motivated by the heuristic considerations in \S\ref{sec1}, and in analogy with
the Theta lift representation of $\log||\Psi_{10}||$ reviewed in \S\ref{sec_psi10}, 
we consider the modular integral
\be
\label{phithetalift}
\tilde\varphi(\Omega)= -\frac12 \int_{\cF_1} \frac{\de^2\tau}{\tau_2^2}\,
\left[ \Gamma^{\rm even}_{3,2}(\Omega;\tau)\, D_{\tau} \tilde h_0(\tau) +  \Gamma^{\rm odd}_{3,2}(\Omega;\tau)\, D_{\tau} \tilde h_1(\tau) \right]\ ,
\ee
where $D_\tau=\frac{\I}{\pi}(\partial_\tau-\frac{\I w}{2\tau_2})$ is the raising operator,
mapping modular forms of weight $w$ to modular forms of weight $w+2$, and
$(\tilde h^0,\tilde h^1)$ are weight $w=-5/2$ modular forms of $\Gamma_0(4)$, associated to the weak Jacobi form $\tilde\phi=\theta^2(\tau,z)/\eta^6(\tau)$ of weight $-2$ and index $1$ in the same way as before:
\be
\label{tildephitheta}
\begin{split}
\tilde\phi(\tau,z) =& \tilde h_0(\tau) \, \theta_{3}(2\tau,2z) +\tilde h_1(\tau) \, \theta_{2}(2\tau,2z) \\
\tilde h_0(\tau)=&\frac{\theta_2(2\tau)}{\eta^6}=2+12 q+56 q^2+\dots\ ,\quad \\
\tilde h_1(\tau)=&-\frac{\theta_3(2\tau)}{\eta^6}=-q^{-1/4} (1+8q+39 q^2+\dots)\ .
\end{split}
\ee
The corresponding weight $-5/2$ modular form in Kohnen's plus space is 
\be
\begin{split}
\tilde h(\tau) = & \tilde h_0(4\tau)+\tilde h_1(4\tau) = 
 -\frac{\theta_4(2\tau)}{[\eta(4\tau)]^6}
= \frac{
\theta_2^8(2\tau) -4 X_2^2(2\tau)}{4 \theta_3(2\tau) \Delta_6}
\\
=& \sum_{m\geq -1} \tilde c(m) q^m = - \left( \frac{1}{q} -2 + 8 q^3 -12 q^4+39 q^7+\dots\right)
\end{split}
\ee
As in the previous case,  $\tilde h(\tau)$ has a simple pole at the cusp at infinity and is regular at the other cusps $\tau=0$ and $\tau=\frac12$. The action of the raising operator $D_\tau$  on  $\tilde h_i$ evaluates to
\be
D_\tau \tilde h_i = \frac{5}{12} \hat E_2\, \tilde h_i - \frac{1}{24} h_i\  ,\quad i=0,1,
\ee
where $\hat E_2= E_2-\frac{3}{\pi \tau_2}$ is the almost holomorphic Eisenstein series of weight 2
and $h_i$ is defined in \eqref{defhi}.
The zero-th Fourier coefficient term of $D_\tau\tilde h_0$ is $-5/(2\pi\tau_2)$, so the integral \eqref{phithetalift} is convergent, with no need for regularization. Using \eqref{DelNarain} and the fact that $D_\tau \tilde h_i$ is an eigenmode of 
$\Delta_{SL(2),-1/2}$ with eigenvalue $5/2$, 
one easily checks that $\tilde\varphi$ is an eigenmode
of $\Delta_{Sp(4)}$ with eigenvalue $5$, away from the separating degeneration divisor $v=0$.
In the limit $v\to 0$, the integrand becomes
\be
\label{g23v}
\Gamma_{2,2} \left[ \theta_3(2\tau)\, D_\tau\tilde h_0(\tau) + \theta_2(2\tau)\, D_\tau\tilde h_1(\tau) \right] 
= -  \Gamma_{2,2}  \stackrel{\tau_2\to\infty}{\sim} -\tau_2\ ,
\ee
leading to a logarithmic divergence. Keeping $v$ small but non zero, retaining the contributions from $m_i=n^i=0, b=\pm 1$  and using $D_\tau \tilde h_1 \sim -\frac12 q^{-1/4} (1-\frac{5}{2\pi\tau_2})$,  we have
\be
\tilde\varphi(\Omega) \sim 
\frac12 \int_1^{\infty} \frac{\de\tau_2}{\tau_2} e^{-\frac{\pi\tau_2|v|^2}{\rho_2\sigma_2-v_2^2}} \left(1-\frac{5}{2\pi\tau_2} \right)
= \frac12 (1+\frac52 z)\, \Gamma(0,\pi z) - \frac{5}{4\pi} e^{-\pi z}\ ,
\ee
where $z=\frac{|v|^2}{\rho_2\sigma_2-v_2^2}$. Thus, in the separating degeneration, we have, in agreement with \eqref{kzsepdeg},
\be
\label{kzsepdeg2}
\tilde\varphi(\Omega) = -\frac12 \left( 1+ \frac{5|v|^2}{2(\rho_2\sigma_2-v_2^2)} \right) \, \log |v|^2 -\log|2\pi
\eta^2(\rho) \eta^2(\sigma)| +\cO(|v|^2)\ ,
\ee
up to terms vanishing analytically as $v\to 0$ (the last term on the right-hand side follows from \eqref{DKL}). The logarithmic singularity implies that 
\be
\label{kzlapp}
\left[ \Delta_{Sp(4)} -5 \right]\, \tilde\varphi = -2\pi\,\det(\Omega_2)\, \delta^{(2)}(v)\ ,
\ee
so $\tilde\varphi$ satisfies the same equation \eqref{kzlap} as $\varphi$. In the remainder of this subsection we extract the asymptotics of  $\tilde\varphi$ 
in the minimal and maximal non-separating degenerations, and find that they agree with the asymptotics of $\varphi$.

\subsubsection{Maximal non-separating degeneration\label{sec_maxnonsep}}

The  maximal  degeneration $\Omega_2\to\infty$ corresponds, in string theory parlance,  to the limit where one of the circles in the torus $T^2$ becomes infinitely large (see footnote \ref{footKawai}). In this limit, the lattice partition function $\Gamma_{3,2}^{\rm even \vert odd }(\Omega)$ factorizes into $\Gamma_{1,1}(r;\tau)\times$ $\Gamma_{2,1}^{\rm even \vert odd }(\tilde\tau;\tau)$, where $r=\sqrt{\det\Omega_2}$ parametrizes the radius of the large circle, and $\tilde\tau=u_2+\I \sqrt{t/\rho_2}$ $\equiv Y+\I R$ the radius $R$ and Wilson line $Y$ for the circle of finite size. It is useful to express 
the lattice partition functions  $\Gamma_{1,1}$ and $\Gamma_{2,1}$ in the `Lagrangian' representation, where modular invariance in $\tau$ is manifest,
\be
\begin{split}
\Gamma_{1,1}(r;\tau)=& r \sum_{(p,q)\in\IZ^2} e^{-r^2 |p+q\tau|^2/\tau_2} \\
\Gamma_{2,1}^{\rm even \vert odd }(\tilde\tau;\tau)=& R\, 
\sum_{\substack{(m,n)\in \IZ^2 \\ b\in 2\IZ \vert 2\IZ+1}}  e^{-\pi R^2 \frac{|m+n\tau|^2}{\tau_2}
+2\I\pi n(m+n\tau) Y^2 + \frac{\I\pi\tau}{2} b^2 +2\I\pi  (m+n\tau) bY} \ .
\end{split}
\ee
In the limit $r\to\infty$, the  $O(3,2,\IZ)=Sp(4,\IZ)$ symmetry is broken to $O(2,1,\IZ)=GL(2,\IZ)$,
acting on
the modulus $\tilde\tau\in\cH_1$ by fractional linear transformations, along with the anti-holomorphic involution $\tilde\tau\mapsto -\bar{\tilde\tau}$.   The leading term in this limit originates from the term $(p,q)=(0,0)$ in 
$\Gamma_{1,1}(r)$, 
\be
\label{phithetalift21}
\varphi_L= -\frac{r}{2} \, \int_{\cF_1} \frac{\de^2\tau}{\tau_2^2}\,
\left[ \Gamma^{\rm even}_{2,1}(\tilde\tau;\tau)\, D_{\tau} \tilde h_0(\tau) +  
\Gamma^{\rm odd}_{2,1}(\tilde\tau;\tau)\, D_{\tau} \tilde h_1(\tau) \right]\ .
\ee
To compute this integral, we decompose the sum over $(m,n)$ in $\Gamma_{2,1}$ into orbits of $SL(2,\IZ)$, obtaining $\varphi_L= \varphi_L^{(0)}+\varphi_L^{(1)}$.  The first term
corresponds to the contribution of the zero orbit $(m,n)=(0,0)$,
\be
\label{phiL0}
\varphi_L^{(0)}= -\frac{r R}{2} \,  \int_{\cF_1} \frac{\de^2\tau}{\tau_2^2}\, \left[ \theta_3(2\tau)\, D_\tau\tilde h_0(\tau) + \theta_2(2\tau)\, D_\tau\tilde h_1(\tau) \right] 
=\frac{\pi r R}{6}\ ,
\ee
since, as already noted in \eqref{g23v}, the term in square bracket reduces to $-1$. The remaining orbits $(m,0)$ with $m\neq 0$ contribute
\be
\varphi_L^{(1)}= -\frac{ r  R}{2}\,  \int_{\cS} \frac{\de^2\tau}{\tau_2^2}\, 
\sum_{m\neq 0} e^{-\pi R^2 m^2/\tau_2} 
\left[ \theta_3(2\tau,2mY)\, D_\tau\tilde h_0(\tau) + \theta_2(2\tau,2mY)\, D_\tau\tilde h_1(\tau) \right] 
\ee
where $\cS$ is the strip $[-1/2,1/2]\times \IR^+$. The integral over $\tau_1$ picks up 
the constant term in $D_\tau\tilde h_0(\tau)$ (corresponding to $b=0$) and the polar
term in $D_\tau\tilde h_1(\tau)$ (corresponding to $b=\pm 1$). Thus we have
 \be
\label{phiL1}
 \begin{split}
\varphi_L^{(1)}=& -\frac{r \, R}{2}\,  \int_{0}^{\infty} \frac{\de\tau_2}{\tau_2^2}\,
 \sum_{m\neq 0} e^{-\pi R^2 m^2/\tau_2}  \\
 &
\times  \left[ \frac{5}{12}\left(1-\frac{3}{\pi \tau_2}\right)\, \left(2-e^{2\pi\I m Y}-e^{-2\pi\I m Y}\right)
 -\frac{1}{24} \left(20+2\, e^{2\pi\I m Y}+2\, e^{-2\pi\I m Y}\right) \right] \\
 =& -\frac{r}{2} \sum_{m=1}^{\infty} \left[ 5 \frac{\cos (2\pi m Y)-1}{\pi^3 R^3 m^4}
 -  \frac{2\cos( 2\pi m Y)}{\pi R m^2} \right]
\end{split}
\ee
Using the identity ${\rm Li}_{k}(e^{2\pi\I x})+(-1)^k {\rm Li}_k(e^{-2\pi\I x}) =  -\frac{(2\pi\I)^k}{k!}\, B_k(x)$
for the polylogarithm, valid for $k$ integer, $0<\Re(x)<1$, 
we arrive, in the region $-\frac12<Y<\frac12, R^2+Y^2>1$, at
 \be
\label{varphiL01}
 \varphi_L\equiv \varphi_L^{(0)}+\varphi_L^{(1)} =
  \frac{\pi r}{6}\, \left[ R+ 5\frac{Y^2(|Y|-1)^2}{R^3} + \frac{1-6|Y|+6Y^2}{R} \right] 
\ee
Setting $\rho_2=r/R, u_2=Y, t=rR$, this reproduces the desired behavior \eqref{phi01lim0} in the
maximal separating degeneration limit ! The square bracket in \eqref{varphiL01}
is recognized as the local modular function $\hat A(\tilde\tau)$ in the two-loop 
supergravity computation \cite[Eq.~(3.8)]{Green:2005ba}.


The subleading terms in the maximal non-separating degeneration limit are obtained by restricting
the sum $\Gamma_{1,1}(r)$ to the orbit representatives $(p,0)$ with $p\neq 0$, and 
unfolding on the strip:
\be
\label{phithetalift21mL}
\tilde\varphi-\varphi_{L}= -\frac{r}{2}\, \int_{\cS} \frac{\de^2\tau}{\tau_2^2}\,
\sum_{p\neq 0} e^{-\pi r^2 p^2/\tau_2}
\left[ \Gamma^{\rm even}_{2,1}(R,Y)\, D_{\tau} \tilde h_0(\tau) 
+  \Gamma^{\rm odd}_{2,1}(R,Y)\, D_{\tau} \tilde h_1(\tau) \right]
\ee
Up to terms of order $e^{-r}$, one can replace $\Gamma^{\rm even}_{2,1}\to \sqrt{\tau_2}$,
$\Gamma^{\rm odd}_{2,1}\to 0$
and $D_{\tau} \tilde h_0(\tau)$ by its constant term $-5/(2\pi\tau_2)$, leading to the next-to-leading correction
\be
\label{phithetalift21NL}
\varphi_{NL}= \frac{15r}{12\pi}\, \int_{0}^{\infty} \frac{\de\tau_2}{\tau_2^{5/2}}\,
\sum_{p\neq 0} e^{-\pi r^2 p^2/\tau_2} = \frac{5 \zeta(3)}{4\pi^2 r^2}\ ,
\ee
consistently with \eqref{E2starexp}. The exponentially suppressed contributions to
$\varphi-\varphi_{L}-\varphi_{NL}$ will be obtained in the analysis of the minimal non-separating
degeneration, to which we now turn.

\subsubsection{Minimal non-separating degeneration\label{sec_minnonsep}}

The limit $\sigma_2\to\infty$ keeping other entries of $\Omega$ fixed corresponds, in string theory parlance,  to the limit  where the volume of the torus $T^2$ becomes infinite, keeping the complex structure 
$\rho$ and holonomy $v$ fixed (see footnote \ref{footKawai}). The Siegel modular group $\Gamma$ is now broken to the Jacobi subgroup $\Gamma_J$. Following \cite{Harvey:1995fq,0919.11036}, the Fourier-Jacobi coefficients (i.e. the Fourier coefficients with respect to $\sigma_1$) can be extracted by applying the orbit method to the lattice partition function written  in the `Lagrangian' representation,
obtained from  \eqref{defGamma32} by Poisson resummation in $m_1,m_2$,
\begin{equation}
\varGamma_{3,2}^{\rm even \vert odd} (\Omega;\tau) =  t \sum_{\substack{A\in \IZ^{2\times 2}\\
b \in 2\IZ  \vert 2\IZ+1}} e^{\frac{\I\pi\tau}{2} b^2 - \pi G(A)}\,,
\end{equation}
where
\begin{equation}
\begin{split}
G(A)  &= \frac{t |\cA|^2}{\rho_2 \, \tau_2}  + 2 \I \sigma \det A + \frac{1}{\rho_2}b\cdot  
( \bar v\, \cA -   v \, \tilde \cA ) 
+\frac{n_2}{\rho_2} ( v^2\, \tilde \cA - \bar v^2 \,\cA ) -  2\I \frac{v_2^2}{\rho_2^2} (n_1 + n_2 \bar \rho ) \, \cA
\end{split}
\end{equation}
and 
\begin{equation}
A = \begin{pmatrix} n_1 & m_1 \\ n_2 & m_2 \end{pmatrix}\,, \qquad
\cA = \begin{pmatrix} 1 & \rho \end{pmatrix} \, A \, \begin{pmatrix} \tau \\ 1 \end{pmatrix}\,, \qquad
\tilde\cA = \begin{pmatrix} 1 & \bar \rho \end{pmatrix} \, A \, \begin{pmatrix} \tau \\ 1 \end{pmatrix}\,.
\end{equation}

The integer matrix $A$ transforms linearly under $SL(2,\IZ)$, and belongs to one of three different types of orbits. 
The orbit $A=0$ produces, as in \eqref{phiL0},
\be
\label{phiL0b}
\varphi^{(0)}= -\frac12 t\,  \int_{\cF_1} \frac{\de^2\tau}{\tau_2^2}\, \left[ \theta_3(2\tau)\, D_\tau\tilde h_0(\tau) + \theta_2(2\tau)\, D_\tau\tilde h_1(\tau) \right] 
=\frac{\pi t}{6}\ .
\ee
The degenerate orbits with ${\rm rk}A=1$  give  instead, restricting to the orbit representatives $(n_1,n_2)=(0,0)$, $(m_1,m_2)\neq (0,0)$ and unfolding on the strip,
\be
\begin{split}
\varphi^{(1)}=&  -\frac12 t \,  \int_{\cS} \frac{\de^2\tau}{\tau_2^2}\, 
\sum_{(m_1,m_2)\neq 0} e^{-\frac{\pi t |m_1+m_2 \rho|^2}{\tau_2 \rho_2} } \\
&\times
\left[ \theta_3(2\tau,2m_1u_2+2 m_2 u_1)\, D_\tau\tilde h_0 + \theta_2(2\tau,2m_1u_2+2 m_2 u_1)\, D_\tau\tilde h_1 \right] \ .
\end{split}
\ee
As in \eqref{phiL1}, the integral over $\tau_1$ picks up 
the constant term in $D_\tau\tilde h_0$ (corresponding to $b=0$) and the polar
term in $D_\tau\tilde h_1$ (corresponding to $b=\pm 1$), leading to 
\be
\label{phiL1b}
 \begin{split}
\varphi^{(1)}=& -\frac{t}{2}\!\!\!
\sum_{(m_1,m_2)\neq 0} \left[ 5 \frac{\rho _2^2(\cos[ 2\pi (m_1u_2+m_2 u_1)]-1)}{\pi^3 t^2 |m_1+m_2 \rho |^4}
 -  \frac{2\rho _2\cos[ 2\pi (m_1u_2+ m_2 u_1)]}{\pi t |m_1+m_2 \rho |^2} \right] \\ &
= \frac12 \cD_{1,1}(\rho ,v) + t^{-1} \left[ \frac{5}{16\pi^2 \rho _2} \cD_{2,2}(\rho ,v) + \frac{5}{2\pi} E^\star(2;\rho )\right]\ .
\end{split}
\ee
Combining \eqref{phiL0b} and \eqref{phiL1b}, we reproduce the desired behavior  \eqref{kzmindeg2}
in the minimal degeneration
limit, with $\varphi_0$ and $\varphi_1$ given in \eqref{phi0D11} and \eqref{phi1D22} !

For the non-degenerate orbits with $\det A\neq 0$, the integral can be unfolded on (a double cover of) the upper 
half-plane, at the expense of restricting the sum to $n_2=0\leq m_1<n_1\neq 0$. Substituting the Fourier series of $\tilde h_0$ and $\tilde h_1$,  the integral over $\tau_1$ is Gaussian, while the integral over $\tau_2$ is of Bessel type. After some algebra (see e.g. \cite[A.2]{Harvey:1995fq}), we arrive at
\be
\label{phiL2b}
 \begin{split}
\varphi^{(2)}= \sum_{\substack{k>0, \ell\geq 0 \\ b\in\IZ}} \tilde c(4k \ell-b^2)\,
\left[ -\frac{5}{16\pi^2 t \rho_2} D_{2,2}\left( x\right)
+\frac12 (4k\ell-b^2) D_{1,1} \left( x \right) \right]\ ,
\end{split}
\ee
where $x=e^{2\pi\I(k\sigma+\ell \rho+b v)}$, and $D_{a,b}(x)$ are the single-valued polylogarithms
defined in \eqref{eq:Dabdef}, 
\be
\label{defd1122}
D_{1,1}(x) = 2\Re[\Li_1(x)]\ ,\quad 
D_{2,2}(x)= -4 \Re[\Li_3(x) -  \log |x|\, \Li_2(x)]\ .
\ee
The formula \eqref{phiL2b} holds in the chamber where $k\sigma_2+\ell \rho_2+b v_2>0$ for all stated values of $(k,\ell, b)$. The sum converges absolutely in a neighborhood of the zero-dimensional cusp $\Omega=\I\infty$ by the same arguments as in \cite{0919.11036,1004.11021}.

\subsection{The Kawazumi-Zhang and Faltings invariants as Theta lifts\label{sec_thm}}

Using the results in \S\ref{sec_kztheta}, \S\ref{sec_minnonsep} and the facts summarized in the introduction, we see that $\hat\varphi=\varphi-\tilde\varphi$ is annihilated by $\Delta_{Sp(4)}-5$ and vanishes up to order $\cO(1/t)$ in the non-separating degeneration limit $t\to\infty$, and up to order  $\cO(|v|^2\log|v|)$ near the separating divisor. On the truncated fundamental domain $\cF_2^{\Lambda}=\cF_2 \cap \{ t < \Lambda, |v|>1/\Lambda\}$, where $\cF_2$ is the standard fundamental domain from \cite{zbMATH03144647}, one has 
\be
\int_{\cF_2^{\Lambda} } \hat\varphi^2 (\star 1) = \frac15 \int_{\cF_2^{\Lambda} } \hat\varphi\,  \Delta_{Sp(4)} \hat\varphi
            = -\frac15 \int_{\cF_2^{\Lambda} } \de\hat\varphi \star \de\hat\varphi  
            +\frac15 \int_{\partial \cF_2^{\Lambda}} \hat\varphi \star \de\hat\varphi\ ,
\ee
where $\star$ denotes the Hodge star on $\cH_2$, and $\partial \cF_2^{\Lambda}$  the boundary of $\cF_2^{\Lambda}$. By the above estimates, the boundary term vanishes in the limit $\Lambda\to\infty$, while the first term converges to a finite, non-positive value. Since the left-hand side is non-negative,  it follows that $\hat\varphi$ must vanish. Thus,
we have shown the

\begin{theorem}
{\it The Kawazumi-Zhang invariant $\varphi(\Omega)$ for compact genus-two Riemann surfaces admits the Theta lift representation
\be
\label{phithetalift2}
\varphi(\Omega)= -\frac12 \int_{\cF_1} \frac{\de^2\tau}{\tau_2^2}\,
\left[ \Gamma^{\rm even}_{3,2}(\Omega;\tau)\, D_{\tau} \tilde h_0(\tau) +  \Gamma^{\rm odd}_{3,2}(\Omega;\tau)\, D_{\tau} \tilde h_1(\tau) \right]\ ,
\ee
where $\Gamma^{\rm even|odd}_{3,2}(\Omega;\tau)$ are the Siegel-Narain theta series defined in \eqref{defGamma32}, and $(\tilde h_0,\tilde h_1)$ is the weight $-\tfrac52$ vector-valued modular form
appearing in the theta series decomposition \eqref{tildephitheta} of the  weak Jacobi form $\theta^2(\tau,z)/\eta^6$ of weight 2 and index 1.}
\end{theorem}

\begin{corollary}
{\it $\varphi(\Omega)$  satisfies the following improved asymptotics: in the minimal non-separating degeneration $t\to+\infty$,
\be
\label{kzmindeg22}
 \varphi(\Omega) =  \frac{\pi}{6} t + \varphi_0 + \frac{\varphi_1}{t} + \cO(e^{-t})\ ,
\ee
where $\varphi_0$ and $\varphi_1$ are defined in  \eqref{phi1D22};
in the maximal non-separating degeneration $L_i\to+\infty$,
\be
\label{kzmaxdeg22}
\varphi(\Omega) = \frac{\pi}{6}
\left[  L_1+L_2+L_3 - \frac{5\, L_1 L_2 L_3}{L_1 L_2+L_2 L_3+L_3 L_1} \right] + 
 \frac{5\zeta(3)}{4\pi^2 \det\Omega_2}+ \cO(e^{-L_i})\ ;
\ee
in the separating degeneration $v\to 0$,
\be
\label{kzsepdeg22}
\varphi(\Omega) = -\frac12 \left( 1+ \frac{5|v|^2}{2(\rho_2\sigma_2-v_2^2)} \right) \, \log |v|^2 -\log|2\pi
\eta^2(\rho) \eta^2(\sigma)| + \cO(|v|^2)\ .
\ee
}
\end{corollary}

\begin{corollary} {\it $\varphi(\Omega)$ admits the Fourier expansion 
\be
\label{KZfull}
 \begin{split}
\varphi(\Omega)= &
\frac{\pi}{6}(\rho_2+\sigma_2-|v_2|) - \frac{5\pi}{6} \frac{|v_2|(\rho_2-|v_2|)(\sigma_2-|v_2|)}{\det\Omega_2}
+ \frac{5\zeta(3)}{4\pi^2 \det\Omega_2}\\
& -\frac{5}{16\pi^2 \det\Omega_2}  \sum_{\substack{(k,\ell,b)>0}} \tilde c(4k \ell-b^2)\,
 D_{2,2}\left( e^{2\pi\I(k\sigma+\ell \rho+b v)}\right) \\
& +\frac12 \sum_{\substack{(k,\ell,b)>0}} (4k\ell-b^2)\, \tilde c(4k \ell-b^2)\, 
D_{1,1} \left( e^{2\pi\I(k\sigma+\ell \rho+b v)} \right) \ ,
\end{split}
\ee
where  $(k,\ell,b)>0$ was defined below \eqref{logPsiprod}, and $D_{1,1}(x)$, $D_{2,2}(x)$ are given in \eqref{defd1122}. The Fourier expansion is absolutely convergent in a neighborhood of the zero-dimensional cusp $\Omega=\I\infty$.}
\end{corollary}

In  \cite{zbMATH06139356}, a relation between the Kawazumi-Zhang invariant $\varphi(\Sigma)$, the Faltings invariant $\delta(\Sigma)$ and the discriminant $\Delta(\Sigma)$ for hyperelliptic compact Riemann surfaces $\Sigma$ was obtained. At genus two, all compact Riemann surfaces are hyperelliptic, and the discriminant is proportional to the Igusa cusp form $\Psi_{10}$. Corollory 1.8 in \cite{zbMATH06139356} states
\be
\label{kzfalt}
\varphi(\Omega) = -3 \log||\Psi_{10}||(\Omega) -\frac52 \delta_F(\Omega) -40\log 2\pi \ .
\ee
Using \eqref{thetalift}, \eqref{phithetalift2} and \eqref{kzfalt},  we obtain

\begin{corollary}
The Faltings invariant admits the Theta lift representation
\be
\begin{split}
\label{thetaliftd}
\delta_F(\Omega)= & \int_{\cF_1} \frac{\de^2\tau}{\tau_2^2}\,
\left[ \Gamma^{\rm even}_{3,2}(\Omega;\tau)\, 
\frac{2 \hat E_2 \tilde h_0+7 h_0}{24} +  \Gamma^{\rm odd}_{3,2}(\Omega;\tau)\, 
\frac{2 \hat E_2 \tilde h_1+7 h_1}{24} - 6\, \tau_2 \right] \\
&
+ 6 \log \left( \frac{4}{3\sqrt3} e^{1-\gamma_E}\right) -10\log 2\pi\ .
\end{split}
\ee
\end{corollary}

The Fourier expansion of $\delta_F(\Omega)$ is easily obtained by combining \eqref{logPsiprod}, \eqref{KZfull} and \eqref{kzfalt}.

\section{Miscellany\label{sec4}}

\subsection{Numerical applications}

The formulae \eqref{logPsiprod} and \eqref{KZfull} provide an efficient numerical procedure for evaluating the Faltings and Kawazumi-Zhang invariants to arbitrary precision. As an illustration, 
for the curve $y^2+y=x^5$ considered in \cite{MR1065156}, with automorphism group $\IZ_5\times \IZ_2$ and period matrix
\be
\Omega =\begin{pmatrix} -\zeta_5^4 & \zeta_5^2 +1 \\ \zeta_5^2 +1 & \zeta_5^2-\zeta_5^3
\end{pmatrix}\ ,
\ee
we find, truncating the sum at $k,\ell,|b|\leq 15$,
\be
\varphi =      0.53801117620500504861\dots\ ,\quad
\delta_F = -16.6790574451477760445\dots\ ,
\ee
where all displayed digits appear to be stable upon varying the truncation. This is consistent with the value $\delta_F=-16,679\dots$ which follows from the numerical computations in \cite[\S 4.5]{MR1065156}.

For another example, consider the curve $y^2=x^6-1$, with automorphism group $D_6\times \IZ_2$ and period matrix 
\be
\Omega =\begin{pmatrix} \frac{2\I}{\sqrt{3}} & \frac{\I}{\sqrt{3}} \\ \frac{\I}{\sqrt{3}}  & \frac{2\I}{\sqrt{3}} \end{pmatrix}\ .
\ee
Using the same truncation, we find
\be		
\varphi =      0.59291015631443383207\dots\ ,\quad                 
\delta_F=                          -16.3412295821338262636\dots
\ee
Finally, consider the Burnside curve $y^2=x(x^4-1)$, with automorphism group $S_4\times \IZ_2$ and period matrix
\be
\Omega =\begin{pmatrix} -\frac12+\frac{\I}{\sqrt{2}} & \frac12 \\ \frac12 & -\frac12+\frac{\I}{\sqrt{2}}
\end{pmatrix}\ .
\ee
Using the same truncation, we find
\be
\varphi =     0.51986038541995901150\dots\ ,\quad                 
\delta_F=          -16.8264632650009721134\dots\ 
\ee

\subsection{Relation to Gromov-Witten invariants}

We note that the Fourier expansion \eqref{KZfull} is similar to \cite[Eq. (A.44)]{Harvey:1995fq}, where modular integrals of the form $\int_{\cF_1} \frac{\de^2\tau}{\tau_2^2}
\Gamma_{8t+2,2} \hat E_2\, F(\tau)$ were considered. Here, $\Gamma_{8t+2,2}$ is a partition of an even self-dual lattice of signature $(8t+2,2)$ and $F(\tau)$ is a weakly
holomorphic modular form of weight $-4t-2$. In this context, the analogue of the coefficients $\tilde c(4k\ell -b^2)$ were identified as the BPS invariants (also known as Gopakumar-Vafa invariants, and closely related to Gromov-Witten invariants) counting rational curves in a suitable K3-fibered Calabi-Yau threefold with $h_{1,1}=8t+3$. It is therefore natural to ask if the  coefficients $\tilde c(4k\ell -b^2)$ in the Fourier expansion of the Kawazumi-Zhang invariant count rational curves in a suitable Calabi-Yau  threefold with $h_{1,1}=4$. Two examples of threefolds with $h_{1,1}=4$ were studied in \cite{Berglund:1996uy,Kawai:1996te} (see also \cite{LopesCardoso:1996nc}). For the example $X(1,1,2,6,10)_{-372}$ in \cite{Berglund:1996uy,LopesCardoso:1996nc}, the rational curves are counted by the weight $-2$ Jacobi form $-(7 E_4 E_{6,1}+5 E_6 E_{4,1})/(6\,\eta^{24})$. For the example $X(2,2,3,3,10)_{-132}$ in \cite{Kawai:1996te}, they are instead counted by the weight $-2$ Jacobi form $-2 E_4 E_{6,1}/\eta^{24}$. The Jacobi form relevant for the Kawazumi-Zhang invariant is proportional to the difference of these two, $\theta^2/\eta^6 = (E_4 E_{6,1}- E_6 E_{4,1})/(144\,\eta^{24})$. It is unclear to the author whether the fact that it is a weak Jacobi form (i.e. has $h(\tau)=\cO(1/q)$ rather than $\cO(1/q^4)$ as in the cases studied in  \cite{Berglund:1996uy,Kawai:1996te,LopesCardoso:1996nc}) disqualifies it from counting rational curves.

\subsection{Holomorphic prepotential}
It is known from \cite{Harvey:1995fq} that modular integrals of the form
\be
\label{defI}
\cI = \int_{\cF_1} \frac{\de^2\tau}{\tau_2^2}\,
\left[ \Gamma^{\rm even}_{3,2}(\Omega;\tau)\, H_0(\tau) +  \Gamma^{\rm odd}_{3,2}(\Omega;\tau)\, H_1(\tau)
- c(0,0)\, \tau_2 \right] \ ,
\ee
where $h(\tau)=h_0(4\tau)+h_1(4\tau)=\sum_{m\geq-\kappa,0\leq \ell \leq 1} c(m,\ell) q^m/\tau_2^\ell$ is an almost weakly holomorphic modular form of weight $-1/2$ and depth $1$ under $\Gamma_0(4)$ in Kohnen's plus space, can be expressed as 
\be
\cI = \Re F_0 + \Re \Box_{-2} F_1 - c(0,0) \log\det\Omega_2\ ,
\ee
where $F_0$ and $F_1$ are holomorphic functions of $\Omega$.  Here,  $\Box_{w}$ is the  Siegel-Maass raising operator
\be
\Box_{w} = -\frac{1}{\pi^2}\left[ \pa_\rho \pa_\sigma - \tfrac14 \pa^2_{v} + 
\frac{\I(1-2 w)}{4(\rho_2\, \sigma_2- v_2^2)} \left( \frac{w}{2\I} + \sigma_2 \pa_\sigma+\rho_2 \pa_\rho 
+ v_2 \pa_{v}\right)  \right]\ ,
\ee
which maps Siegel modular forms of weight $w$ to modular forms of weight $w+2$. $F_0$ is the logarithm of a holomorphic Siegel modular form of weight $-2c(0,0)$. $F_1$, known as the holomorphic prepotential,
is ambiguous modulo elements in the kernel $\cK$ of the operator $\Re(\Box_{-2})$. The latter
includes  quadratic polynomials in $(\rho,\sigma,v)$ with imaginary coefficients, as well as cubic polynomials of the form $(\rho\sigma-v^2) (\alpha \rho+\beta \sigma)$ where $\alpha,\beta$ are imaginary.  Since the integral $\cI$ is a Siegel modular function, $F_1$ must transform under $\gamma\in Sp(4,\IZ)$ as 
\be
F_1\vert_{-2} \gamma \, (\Omega) = F_1(\Omega) + P_\gamma(\Omega)\ ,
\ee
where $P_\gamma(\Omega)$ is an element in $\cK$. Thus, $F_1$ is a mock-type holomorphic Siegel modular form of weight $-2$.  
For the modular integral \eqref{thetalift}, the modular form $H(\tau)=-\tfrac14 h(\tau)$ is weakly holomorphic therefore $F_1$ vanishes, while  $F_0=\log\Psi_{10}$ (up to an additive constant). 
For the modular integral \eqref{phithetalift}, $H(\tau)=-\frac18 D_\tau\tilde h(\tau)$ is 
the modular derivative of a weakly holomorphic form, therefore $F_0$ vanishes \cite{Angelantonj:2015rxa,AFP5-to-appear}. Since, by Theorem 1, $\varphi(\Omega)$ is equal to $\cI$ for this choice of $H(\tau)$, we have the
 
 \begin{corollary} 
 {\it The Kawazumi-Zhang invariant $\varphi(\Omega)$ is equal to the real part of the action of the Siegel-Maass raising operator $\Box_{-2}$ on the `holomorphic prepotential' $F_1(\Omega)$,
\be
\label{phiBoxF}
\varphi= \Re\left(  \Box_{-2} F_1 \right)
\ee 
where 
\be
\label{prepF}
F_1(\Omega) = \sum_{(k,\ell,b)>0} \tilde c(4k\ell-b^2)\, {\rm Li}_3\left( e^{2\pi \I(k \sigma+\ell \rho+b v)} \right) 
-\frac{\I \pi^3}{3} \rho\sigma(\rho+\sigma-2v) + \zeta(3) \ .
\ee
}
\end{corollary}

{\noindent \it Proof}: Using \cite[A.32]{Kiritsis:1997hf} 
\be
\begin{split}
\square^{n}_{-2n} {\rm Li}_{2n+1}(x) =& \sum_{r=0}^{n}
\frac{n! (n+r-w)!}{r! (n-r)!\, (n-w)!} \frac{2^{2n-2r}\, (k\ell-b^2)^{n-r} }{(\pi \det\Omega_2)^r}
\, L_{(r)}\left(\frac{\log{x}}{2\pi \I}\right) \\
\end{split}
\ee
with $n=2$, where $x=e^{2\pi\I(k\sigma+\ell \rho+b v)}$ and
\be
\label{CombinedPolylog}
	L_{(r)}(z)=\sum_{m=0}^{r}\frac{(r+m)!}{m!(r-m)!(4\pi)^m}[\Im z]^{r-m}\,\textrm{Li}_{r+m+1}(e^{2\pi i z}) 
\ee
is related to the Bloch-Wigner-Ramakrishnan polylogarithm $D_{a,b}$ in \eqref{eq:Dabdef} 
via \cite{AFP5-to-appear}
\be
D_{r+1,r+1}(x)  =  2\Re \left[ \frac{(-4\pi)^r}{r!} \, L_{(r)}\left(\frac{\log{x}}{2\pi \I}\right) \right]\ ,
\ee
one easily checks that the action of $\Re(\Box_{-2})$ on the first term
of \eqref{prepF} produces the last two lines in \eqref{KZfull}. The action of the same on the polynomial terms
in \eqref{prepF} produces the first line in \eqref{KZfull}. \hfill $\square$

\medskip

\noindent {\it Remark.} More generally, integrals of the form \eqref{defI}, where $H(\tau)$ is an almost weakly holomorphic modular form of weight $-1/2$ and depth $n$, can be expressed as 
\be
\cI = \sum_{r=0}^n \Re \Box_{-2r} F_r - c(0,0) \log\det\Omega_2\ ,
\ee
where $F_r$ are holomorphic functions of $\Omega$ known as generalized prepotentials, which transform as mock-type Siegel modular forms of weight $-2r$ \cite{Kiritsis:1997hf,Angelantonj:2015rxa,AFP5-to-appear}. When $H(\tau)$ is obtained by acting $r$ times with the raising operator $D_\tau$ on a holomorphic modular form $h(\tau)$ of weight $-2r-\tfrac12$, then all $F_r$ vanish
except $F_n$ \cite{Angelantonj:2015rxa,AFP5-to-appear}.

\subsection{Quartic differential equation}

Observe that the Narain partition function satisfies, in addition to \eqref{DelNarain},
\be
\left[ \underline{\Box}_2\, \Box_0 - \frac{1}{16\pi^4} \Delta_{SL(2),1/2}\, 
\left(\Delta_{SL(2),1/2}-\frac{1}{2}\right) 
\right]\, \Gamma_{3,2}^{\rm even|odd} = 0\ ,
\ee
where  $ \underline\Box_w$ is the Siegel-Mass lowering operator (formally independent of $w$)
\be
 \underline\Box_w= 
 -\pi^2(\rho_2\sigma_2-v_2^2)^2\, \left[ \pa_{\bar\rho} \pa_{\bar\sigma} - \tfrac14 \pa^2_{\bar v} - 
\frac{\I}{4(\rho_2\, \sigma_2- v_2^2)} \left( \sigma_2 \pa_{\bar\sigma}+\rho_2 \pa_{\bar\rho} 
+ v_2 \pa_{\bar v}\right)  \right]  \ ,
\ee
which maps Siegel modular forms of weight $w$ to Siegel modular forms of weight $w-2$. 
By integration by parts, we conclude that

\begin{corollary} 
{\it $\varphi(\Omega)$ satisfies the quartic differential equation (away from the separating degeneration) }
\be
\label{quarticphi}
\left( \underline\Box_2\, \Box_0-\frac{15}{32} \right) \varphi = 0\ .
\ee
\end{corollary}

It would be interesting to understand  the fate of the differential equations \eqref{kzlap} and \eqref{quarticphi} at higher genus.

\acknowl{ I am very grateful to Rodolfo Russo  for collaboration at an early stage of this work. In addition, I wish to thank Eric d'Hoker, Michael Green  and Rodolfo Russo
for collaboration on \cite{D'Hoker:2014gfa}, Robin de Jong and Stephen Miller for useful comments on an earlier version of this work,  and  Carlo Angelantonj and Ioannis Florakis for an on-going collaboration on generalized Borcherds lifts, which paved the way for the present work. I am also grateful to the organizers of the workshop "Automorphic Forms: Advances and Application" (CIRM, Marseille, May 25-29, 2015) for the opportunity to present this work, and to the workshop participants for stimulating discussions.}


\begin{thebibliography}{10}
\providecommand{\url}[1]{\texttt{#1}}
\providecommand{\urlprefix}{URL }
\expandafter\ifx\csname urlstyle\endcsname\relax
  \providecommand{\doi}[1]{doi:\discretionary{}{}{}#1}\else
  \providecommand{\doi}{doi:\discretionary{}{}{}\begingroup
  \urlstyle{rm}\Url}\fi

\bibitem{AlvarezGaume:1987vm}
\textit{L.~Alvarez-Gaume}, \textit{J.~Bost}, \textit{G.~W. Moore},
  \textit{P.~C. Nelson} and \textit{C.~Vafa}, {Bosonization on Higher Genus
  Riemann Surfaces}, Commun.Math.Phys. \textbf{112} (1987), 503.
  \doi{10.1007/BF01218489}.

\bibitem{AFP5-to-appear}
\textit{C.~Angelantonj}, \textit{I.~Florakis} and \textit{B.~Pioline}, in preparation.
 

\bibitem{Angelantonj:2012gw}
\textit{C.~Angelantonj}, \textit{I.~Florakis} and \textit{B.~Pioline},
  {One-Loop BPS amplitudes as BPS-state sums}, JHEP \textbf{1206} (2012), 070.
  \doi{10.1007/JHEP06(2012)070}

\bibitem{Angelantonj:2015rxa}
\textit{C.~Angelantonj}, \textit{I.~Florakis} and \textit{B.~Pioline},
  {Threshold corrections, generalised prepotentials and Eichler integrals},
  Nucl.Phys. \textbf{B897} (2015), 781--820.
  \doi{10.1016/j.nuclphysb.2015.06.009}

\bibitem{Antoniadis:1995ct}
\textit{I.~Antoniadis}, \textit{S.~Ferrara}, \textit{E.~Gava},
  \textit{K.~Narain} and \textit{T.~Taylor}, {Perturbative prepotential and
  monodromies in N=2 heterotic superstring}, Nucl.Phys. \textbf{B447} (1995),
  35--61. \doi{10.1016/0550-3213(95)00240-S}

\bibitem{Berglund:1996uy}
\textit{P.~Berglund}, \textit{S.~H. Katz}, \textit{A.~Klemm} and
  \textit{P.~Mayr}, {New Higgs transitions between dual N=2 string models},
  Nucl.Phys. \textbf{B483} (1997), 209--228. 
  \doi{10.1016/S0550-3213(96)00450-6}

\bibitem{0919.11036}
\textit{R.~E. Borcherds}, {Automorphic forms with singularities on
  Grassmannians.}, Invent. Math. \textbf{132} (1998), no.~3, 491--562.
  \doi{10.1007/s002220050232}

\bibitem{MR1065156}
\textit{J.-B. Bost}, \textit{J.-F. Mestre} and \textit{L.~Moret-Bailly}, Sur le
  calcul explicite des ``classes de {C}hern'' des surfaces arithm\'etiques de
  genre {$2$}, Ast\'erisque  (1990), no. 183, 69--105. S{\'e}minaire sur les
  Pinceaux de Courbes Elliptiques (Paris, 1988).

\bibitem{1004.11021}
\textit{J.~H. Bruinier}, {Borcherds products on $O(2,l)$ and Chern classes of
  Heegner divisors}, Lecture Notes in Mathematics {\bf 1870}, Springer, 2002.

\bibitem{zbMATH05725877}
\textit{R.~{de Jong}}, {Admissible constants for genus 2 curves.}, {Bull. Lond.
  Math. Soc.} \textbf{42} (2010), no.~3, 405--411. \doi{10.1112/blms/bdp132}

\bibitem{zbMATH06139356}
\textit{R.~{de Jong}}, {Second variation of Zhang's $\lambda $-invariant on the
  moduli space of curves.}, {Am. J. Math.} \textbf{135} (2013), no.~1,
  275--290. \doi{10.1353/ajm.2013.0008}

\bibitem{zbMATH06355718}
\textit{R.~{De Jong}}, {Asymptotic behavior of the Kawazumi-Zhang invariant for
  degenerating Riemann surfaces.}, {Asian J. Math.} \textbf{18} (2014), no.~3,
  507--524.

\bibitem{zbMATH06339620}
\textit{E.~{D'Hoker}} and \textit{M.~B. {Green}}, {Zhang-Kawazumi invariants
  and superstring amplitudes.}, {J. Number Theory} \textbf{144} (2014),
  111--150. \doi{10.1016/j.jnt.2014.03.021}

\bibitem{D'Hoker:2014gfa}
\textit{E.~D'Hoker}, \textit{M.~B. Green}, \textit{B.~Pioline} and
  \textit{R.~Russo}, {Matching the $D^{6}R^{4}$ interaction at two-loops}, JHEP
  \textbf{1501} (2015), 031. \doi{10.1007/JHEP01(2015)031}

\bibitem{Dixon:1990pc}
\textit{L.~J. Dixon}, \textit{V.~Kaplunovsky} and \textit{J.~Louis}, Moduli
  dependence of string loop corrections to gauge coupling constants, Nucl.
  Phys. \textbf{B355} (1991), 649--688.

\bibitem{MR781735}
\textit{M.~Eichler} and \textit{D.~Zagier}, The theory of {J}acobi forms,
  \textit{Progress in Mathematics}, volume~55, Birkh\"auser,
  Boston, 1985.

\bibitem{zbMATH03891504}
\textit{G.~{Faltings}}, {Calculus on arithmetic surfaces.}, {Ann. Math. (2)}
  \textbf{119} (1984), 387--424. \doi{10.2307/2007043}

\bibitem{zbMATH03144647}
\textit{E.~{Gottschling}}, {Explizite Bestimmung der Randfl\"achen des
  Fundamentalbereiches der Modulgruppe zweiten Grades.}, {Math. Ann.}
  \textbf{138} (1959), 103--124.  \doi{10.1007/BF01342938}

\bibitem{Green:2008bf}
\textit{M.~B. Green}, \textit{J.~G. Russo} and \textit{P.~Vanhove}, {Modular
  properties of two-loop maximal supergravity and connections with string
  theory}, JHEP \textbf{0807} (2008), 126. \doi{10.1088/1126-6708/2008/07/126}

\bibitem{Green:2005ba}
\textit{M.~B. Green} and \textit{P.~Vanhove}, {Duality and higher derivative
  terms in M theory}, JHEP \textbf{0601} (2006), 093.
  \doi{10.1088/1126-6708/2006/01/093}

\bibitem{MR1428063}
\textit{V.~A. Gritsenko} and \textit{V.~V. Nikulin}, Siegel automorphic form
  corrections of some {L}orentzian {K}ac-{M}oody {L}ie algebras, Amer. J. Math.
  \textbf{119} (1997), no.~1, 181--224.

\bibitem{Harvey:1995fq}
\textit{J.~A. Harvey} and \textit{G.~W. Moore}, {Algebras, BPS States, and
  Strings}, Nucl. Phys. \textbf{B463} (1996), 315--368.

\bibitem{Kawai:1995hy}
\textit{T.~Kawai}, {N=2 heterotic string threshold correction, K3 surface and
  generalized Kac-Moody superalgebra}, Phys.Lett. \textbf{B372} (1996), 59--64.
  \doi{10.1016/0370-2693(96)00052-4}

\bibitem{Kawai:1996te}
\textit{T.~Kawai}, {String duality and modular forms}, Phys.Lett. \textbf{B397}
  (1997), 51--62. \doi{10.1016/S0370-2693(97)00146-9}

\bibitem{Kawazumi}
\textit{N.~Kawazumi}, {Johnson's homomorphisms and the Arakelov-Green function}
   (2008).

\bibitem{Kiritsis:1997hf}
\textit{E.~Kiritsis} and \textit{N.~A. Obers}, {Heterotic/type-I duality in $D
  < 10$ dimensions, threshold corrections and D-instantons}, JHEP \textbf{10}
  (1997), 004.

\bibitem{LopesCardoso:1996nc}
\textit{G.~Lopes~Cardoso}, \textit{G.~Curio} and \textit{D.~Lust},
  {Perturbative couplings and modular forms in N=2 string models with a Wilson
  line}, Nucl.Phys. \textbf{B491} (1997), 147--183.
  \doi{10.1016/S0550-3213(97)00047-3}

\bibitem{Mayr:1993mq}
\textit{P.~Mayr} and \textit{S.~Stieberger}, {Threshold corrections to gauge
  couplings in orbifold compactifications}, Nucl. Phys. \textbf{B407} (1993),
  725--748. \doi{10.1016/0550-3213(93)90096-8}

\bibitem{PiolineRusso-to-appear}
\textit{B.~Pioline} and \textit{R.~Russo}, {Infrared divergences and harmonic anomalies in the
two-loop superstring effective action}, JHEP \textbf{1512} (2015) 102.
\doi{10.1007/JHEP12(2015)102}

\bibitem{MR1105425}
\textit{R.~Wentworth}, The asymptotics of the {A}rakelov-{G}reen's function and
  {F}altings' delta invariant, Comm. Math. Phys. \textbf{137} (1991), no.~3,
  427--459. 

\bibitem{zbMATH04144378}
\textit{D.~{Zagier}}, {The Bloch-Wigner-Ramakrishnan polylogarithm function.},
  {Math. Ann.} \textbf{286} (1990), no. 1-3, 613--624.
  \doi{10.1007/BF01453591}

\bibitem{zbMATH05661751}
\textit{S.-W. {Zhang}}, {Gross-Schoen cycles and dualising sheaves.}, {Invent.
  Math.} \textbf{179} (2010), no.~1, 1--73. \doi{10.1007/s00222-009-0209-3}

\end{thebibliography}

\end{document}